\documentclass[a4paper]{jpconf}
\usepackage{graphicx}
\usepackage{amsfonts,amsmath,amssymb,mathrsfs,bm,amsthm}
\begin{document}
\title{Precision Measurements of the Fundamental Properties of the Proton and Antiproton}


\author{C. Smorra}
\address{RIKEN, Fundamental Symmetries Laboratory, 2-1 Hirosawa, Wako, Saitama, 351-0198, Japan}
\address{CERN, Esplanade des Particules 1, 1217 Meyrin, Switzerland}
\ead{christian.smorra@cern.ch}

\author{A. Mooser}
\address{RIKEN, Fundamental Symmetries Laboratory, 2-1 Hirosawa, Wako, Saitama, 351-0198, Japan}
\address{Max-Planck-Institut f{\"u}r Kernphysik, Saupfercheckweg 1, D-69117, Heidelberg, Germany}
\ead{andreas.mooser@mpi-hd.mpg.de}

\begin{abstract}
Precision measurements comparing the fundamental properties of conjugate particles and antiparticles  constitute stringent tests of CPT invariance. 
We review recent precision measurements of the BASE collaboration, which improved the uncertainty of the proton and antiproton magnetic moments and the comparison of the proton-to-antiproton charge-to-mass ratio. 
These measurements constitute the most stringent tests of CPT invariance with antiprotons. 
Further, we discuss the improved limit on the antiproton lifetime based on the storage of a cloud of antiprotons in the unique BASE reservoir trap. 
Based on these recent advances, we discuss ongoing technical developments which comprise a coupling trap for the sympathetic cooling of single (anti-)protons with laser-cooled beryllium ions, a transportable trap to relocate antiproton measurements into a high-precision laboratory, and a new experiment to measure the magnetic moment of helium-3 ions, which will improve absolute precision magnetometry.
\end{abstract}


\section{Introduction}
In the development of the Standard Model of particle physics, symmetries have played an important role \cite{Weinberg2018}. 
In the early last century, the discrete symmetries of charge- (C), parity- (P), and time-reversal (T), and the combined CP-symmetry were believed to be individually conserved. However, experiments showed that, for example, the P-symmetry is violated in the nuclear beta decay \cite{WuExperiment}, and that the combined CP-symmetry is violated in the decay of neutral kaons \cite{CroninKaon}. 
These experimental observations were included into the Standard Model as the V-A theory of the weak interaction and a CP-violating phase in the quark-mixing matrix, respectively.
At present, the interactions in the Standard Model are described by local Lorentz-invariant quantum-field theories.  
These are expected to be invariant under the combined CPT transformation \cite{CPT1}. 
Therefore, CPT invariance is regarded as one of the fundamental symmetries in the Standard Model and should hold exactly.
As consequences of this symmetry, conjugate particle-antiparticle pairs are created and annihilated in pair processes and have identical fundamental properties except for signs.
This theoretical understanding is in conflict to our astronomical observations, which indicate that our Universe consists almost exclusively of matter \cite{BaryonAsymmetry}. 
This indicates that our understanding of the fundamental interactions is incomplete, since the Standard Model can neither explain the matter-antimatter asymmetry in our Universe, nor reproduce the observed matter excess based on the CP violation in the quark sector \cite{BAU-review}. 
Searching for additional sources of symmetry violation or additional interactions may provide important hints to improve our understanding of the matter-antimatter asymmetry.
So far, CPT invariance is holding up to all conducted experimental tests \cite{CPTTables}. 
Therefore, it is essential to make more stringent CPT tests with increased experimental precision or to test CPT invariance in new systems. \\


Experimental tests of CPT invariance with antiprotons are conducted so far at the antiproton decelerator (AD) of CERN \cite{Maury1999HypInt}, which is presently the world's only low-energy antiproton facility. 
Here, high-precision spectroscopy is conducted on antiprotons \cite{JerryAntiproton,JackAntiproton,UlmerNature2015,SmorraNature}, antihydrogen \cite{ALPHA,ALPHA2,ASACUSAHBarBeam,ASACUSAHydrogen}, and antiprotonic helium \cite{Masaki}. 
Opportunities on measurements in other exotic systems are presently discussed, such as on the positively-charged antihydrogen ion $\overline{H}^+$\cite{WalzHbarPlus,GBAR} or the antihydrogen molecular ion $\overline{\mathrm{H}}_2^-$ \cite{DehmeltHbar2Ion,MyersHbar2Ion}.
Of particular interest is also the search for a modified antimatter gravitation \cite{GBAR,AEgIS,ALPHAg,ALPHAGrav}. 
The ongoing experimental efforts will benefit from the new deceleration stage ELENA \cite{ELENA}, which reduces the antiproton energy for the experiments by a factor $\sim$ 50 compared to the AD, down to 100 keV. \\

In the experiments of the BASE collaboration, we compare the fundamental properties of protons and antiprotons, such as their lifetimes $\tau_{p/\overline{p}}$, charge-to-mass ratios $(q/m)_{p/\overline{p}}$, and magnetic moments $\mu_{p/\overline{p}}$. 
The magnetic moment of the proton $\mu_p$ was known with a relative precision of 10 parts per billion (ppb) from a measurement using a hydrogen maser in 1972 \cite{Winkler1972}, which provided the best measurement of this fundamental property for more than four decades. 
The antiproton magnetic moment $\mu_{\overline{p}}$ was until recently only known with a relative uncertainty on the 0.1$\,\%$ level from the spectroscopy of antiprotonic helium atoms \cite{Pask2009}. 
Experimental techniques for high-precision measurements of magnetic moments of single particles with $\sim\,1\,$ppb precision were already developed in the 1970s and 1980s in the concourse of electron and positron measurements by the group of Dehmelt \cite{DehmeltCSG,Monoelectron,Dehmeltg-2,DehmeltCPT}.
These methods provided a promising basis to conduct also high-precision magnetic moment measurements of protons and antiprotons with comparable precision \cite{Quint2004}. Extending the methods for protons and antiprotons required however to considerably advance several experimental techniques, since their magnetic moment is about a factor 660 smaller than those of the leptons. 
Essential steps to realize such measurements for protons and antiprotons were the development of the double-trap technique to suppress line-broadening effects from magnetic field gradients \cite{haeffner2003double}, the development of a Penning-trap system to detect spin-transitions of a single proton at first with a statistical method \cite{UlmerPRL2011}, and to reach the sensitivity of detecting single spin transitions with protons \cite{MooserPRL2013} and antiprotons \cite{SmorraPLB2017}. 
These techniques were implemented in two experiments: the proton $g$-factor experiment at the University of Mainz (BASE-Mainz) \cite{CCRodegheri2012}, and the BASE experiment at the antiproton decelerator of CERN \cite{SmorraEPJST2015}.
These developments resulted in the first direct measurements of the proton and antiproton magnetic moments using a single-trap method with an intermediate level of precision of the order 10$^{-6}$ \cite{JackAntiproton,CCRodegheri2012,JackProton,HiroNC2017}.
The application of the double-trap technique with the sensitivity on detecting single spin-transitions resulted in the first parts-per-billion measurements of the proton and antiproton magnetic moments \cite{SmorraNature,MooserNature}, and recently even in a 0.3$\,$ppb measurement of the proton magnetic moment \cite{SchneiderScience2017}.
The proton measurements constitute the most precise measurement of a nuclear magnetic moment and decrease the uncertainty of the proton magnetic moment by a factor of 33 compared to the previous best measurement \cite{Winkler1972}, and the antiproton measurement improves the uncertainty by more than six orders of magnitude compared to exotic atom spectroscopy \cite{Pask2009}, and a factor of 350 compared to the best single-trap measurement \cite{HiroNC2017}, which was also performed by the BASE collaboration.
By combining these measurements, a 10$^6$-times more stringent test of CPT invariance in the baryon sector was realized and improved limits on CPT-odd antiproton coefficients in the non-minimal Standard Model Extension \cite{Ding2016} and on CPT-odd dimension-five operators \cite{Stadnik2014}, which would cause a splitting of the proton/antiproton magnetic moments were reported.
The developed methods also facilitated improved comparisons of the proton and antiproton charge-to-mass ratio \cite{JerryAntiproton,UlmerNature2015} and improved limits on the antiproton lifetime from laboratory experiments \cite{Sellner2017}. \\

In the future, further improvements on testing CPT invariance in the baryon sector require improved experimental techniques for single protons and antiprotons. To this end, we are developing a sympathetic-cooling method for (anti)protons using a common-endcap electrode trap system \cite{Wineland,Bohman2018}, and the BASE collaboration also targets to implement a quantum-logic method to improve the spin state determination \cite{QLEDS2018}. In addition, we target to lower the ambient magnetic field fluctuations by building a transportable antiproton container to relocate our precision measurements into the low-noise environment of a high-precision laboratory. Based on these recent developments, we are also building an experiment to measure the magnetic moment of $^3$He$^{2+}$ with 1 ppb precision, which requires a 10-fold more-sensitive apparatus compared to (anti)proton magnetic moment measurements. This measurement will allow to establish hyperpolarized helium-3 as independent high-precision magnetometer.


\section{The basic experimental techniques}
Penning traps are presently the tool of choice for high-precision measurements of magnetic moments \cite{SmorraNature,MooserNature,SchneiderScience2017,Hanneke2008,SturmAtoms} and charge-to-mass ratios \cite{UlmerNature2015,HeisseProton,MyersAtoms} of single particles. For these measurements, a superposition of a strong magnetic field $\vec{B}=B_0 \hat{e}_z$ ($B=1.9\,$T) for radial confinement and a weak electric quadrupole field $\Phi = V_0 C_2 (z^2-\rho/2^2)$ ($V_0 \sim 7\,$V) confines a single charged particle, where $C_2$ is a trap specific geometry factor, and $V_0$ the voltage applied to the ring electrode of the trap. 
The trapped particle has three eigenmotions, the axial mode is a harmonic oscillation along the magnetic field lines, and the modified cyclotron mode and magnetron mode comprise the motion in the radial plane \cite{Brown}. 
In this configuration, we make high-precision measurements of the frequencies of single trapped particles.
The magnetic moment of the proton and antiproton are measured by determining the frequency ratio of the Larmor (spin-precession) frequency $\omega_L=(g/2)\,(q/m)\,B$ and the cyclotron frequency $\omega_c= (q/m)\,B$:
\begin{eqnarray}
\frac{\omega_L}{\omega_c}=\frac{g_{p/\overline{p}}}{2}=\pm\frac{\mu_{p/\overline{p}}}{\mu_N},
\end{eqnarray}
where $g_{p/\overline{p}}$ denotes the dimensionless magnetic moment of the proton/antiproton ($g$-factor), and $\mu_N$ the nuclear magneton. 
The cyclotron frequency $\nu_c$ is determined by measuring the three eigenfrequencies of the trapped particle using the relation $\nu_c^2 = \nu_+^2 + \nu_z^2 + \nu_-^2$ \cite{Brown}. 
The Larmor frequency is determined by driving spin-transitions with an oscillating magnetic field and measuring the spin-transition probability as a function of the drive frequency. \\

To determine $\nu_c$, the motional frequencies of the trapped particle are measured non-destructively by using image-current detection methods \cite{Wine,UlmerPRL5Dip,HiroRSI}. To this end, the current resulting from the movement of the image charge induced in the trap electrodes is picked up by a superconducting coil which is attached to one of the trap electrodes. This forms a parallel tank circuit, which acts on its resonance frequency as a large resistor. Matching the axial frequency of the particle to the resonance frequency of the tank circuit converts the image-current signal into a detectable voltage drop, which is read-out using cryogenic low-noise amplifiers and a Fast-Fourier Transform (FFT) analyzer. The detection systems reach single particle sensitivity for (anti)protons in traps with $d < 10\,$mm diameter, and provide measurements of the axial frequency $\nu_z$ with $\sim20\,$mHz resolution in $\sim1\,$min averaging time \cite{HiroRSI}. To measure the cyclotron frequency, we use a sideband method \cite{CornellSB}, where we couple the radial and axial modes and determine the cyclotron frequency from the axial and sideband frequencies \cite{UlmerPRL5Dip}. Using this method, such measurements reach a precision of a few ppb in about 2 minutes averaging time \cite{UlmerNature2015,Higuchi}.

An essential technique in magnetic measurements is to detect spin transitions for the spectroscopy of the Larmor resonance. For this purpose, we utilize the continuous Stern-Gerlach effect and make quantum non-demonlition measurements of the spin state of single trapped particles \cite{DehmeltCSG}. To this end, a strong magnetic bottle $B_z(z,\rho)=B_0 + B_2 (z^2-\rho^2/2)$ is superimposed to one of our traps, the \textit{analysis trap}. The magnetic potential of the bottle couples the orbital and spin angular momentum to the axial mode, such that quantum transitions of the spin or in the radial modes change the axial frequency:
\begin{eqnarray}
\Delta\nu_z=\frac{1}{4 \pi m \nu_z}\frac{B_2}{B_0} h \nu_+ \left(\frac{g}{2} m_s + (n_+ + \frac{1}{2})+\frac{\nu_-}{\nu_+}(n_- + \frac{1}{2})\right),
\label{eq:CSGE}
\end{eqnarray}
where $B_2$ the curvature of the magnetic field, and $n_+$, $n_-$, and $m_s = \pm 1/2$, are the cyclotron, magnetron, and spin quantum numbers, respectively. 
Single quantum transitions, $\Delta n_+$=1, $\Delta n_-$=1, and $\Delta m_s$=1, cause frequency shifts of about 65$\,$mHz, 20$\,\mu$Hz and 180$\,$mHz, respectively.  
Therefore, the challenge for the magnetic moment measurement is to distinguish spin transitions and cyclotron transitions that are driven by residual electric field noise in the trap. 
In the cryogenic 3.6$\,$mm diameter trap of the BASE apparatus at CERN, the electric field noise reaches levels corresponding to a few cyclotron transitions per hour for particles below 50$\,$mK cyclotron energy \cite{BorchertPRL}. 
This enables identifying single spin transitions with high fidelity \cite{MooserPRL2013,SmorraPLB2017}.
A second challenge is the line broadening imposed by the magnetic bottle. Whereas electron $(g-2)$ measurements can be performed in a single trap with a moderate $B_2$ of $150\,$T/m$^2$ \cite{Brown}, the (anti)proton requires to go with the magnetic bottle strength close to the technical limit by using a ferromagnetic ring electrode, which results in $B_2 \sim 300000\,$T/m$^2$. 
This causes a significant line broadening of the spin and cyclotron resonance in this trap and limits the uncertainties of magnetic moment measurements of a few times 10$^{-7}$ \cite{HiroNC2017}. 
To reach a higher level of precision, it is essential to use the double-trap technique, where the spin-state identification takes place in the strongly inhomogeneous analysis trap, and the spectroscopy of the Larmor frequency is spatially separated into a second homogeneous trap, the \textit{precision trap}, where the residual magnetic gradients and their line-broadening effects are at least a factor 10$^5$ smaller. 
Using this technique, the latest measurement of the proton magnetic moment reached down to 0.3 ppb in relative precision \cite{SchneiderScience2017}.


\section{Precision measurement of the proton magnetic moment}

\begin{figure}[h!]
\centering
\includegraphics[width= 0.984 \textwidth]{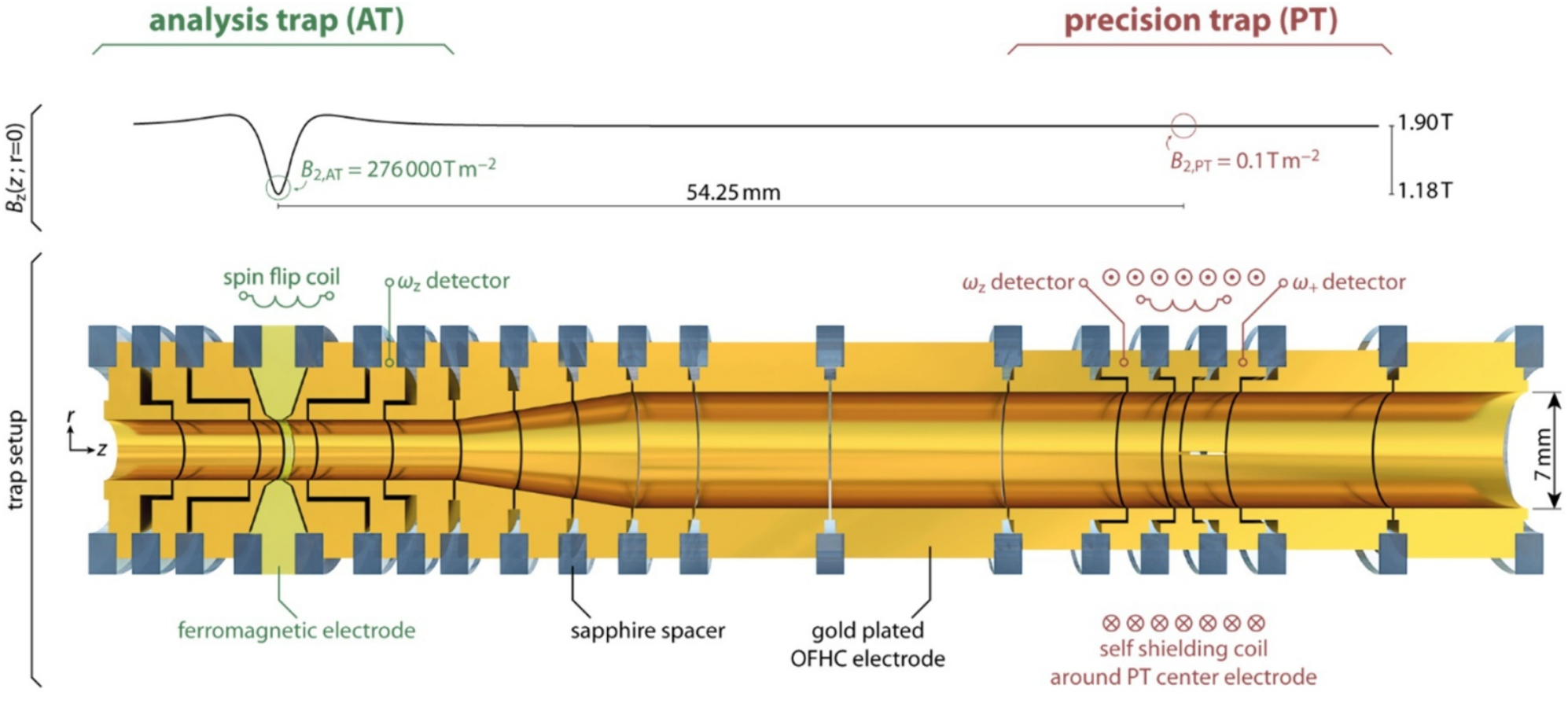}
\caption{\textbf{Double Penning-trap system for the proton magnetic moment measurement.} 
The electrode stack in combination with the magnetic field pointing along the symmetry axis of the electrode form the Penning traps for the proton magnetic moment measurement.
The gold-plated OFE copper electrodes are shown in gold and the sapphire rings providing the insulation in blue.
The spin-state analysis trap has a ferromagnetic ring electrode shown in green to form the strong magnetic bottle required to apply the continuous Stern-Gerlach effect.
The analysis trap and the homogeneous precision trap for high-precision frequency measurements are spatially seperated by several transport electrodes by a distance of 54.25$\,$mm to decrease the residual inhomogeneity due to the magnetic bottle in the precision trap. 
A self-shielding coil stabilizes the magnetic field in the precision trap. 
Both traps have an axial image-currect ($\omega_z$) detector, and spin-flip coil in their vicinity to irradiate oscillating magnetic fields to drive spin flips.
The precision trap has in addition a cyclotron $\omega_+$ detector for resistive cooling of the cyclotron mode. }
\label{fig:ProtonTrap}
\end{figure}

\begin{figure}[h!]
\centering
\includegraphics[width= 0.984 \textwidth]{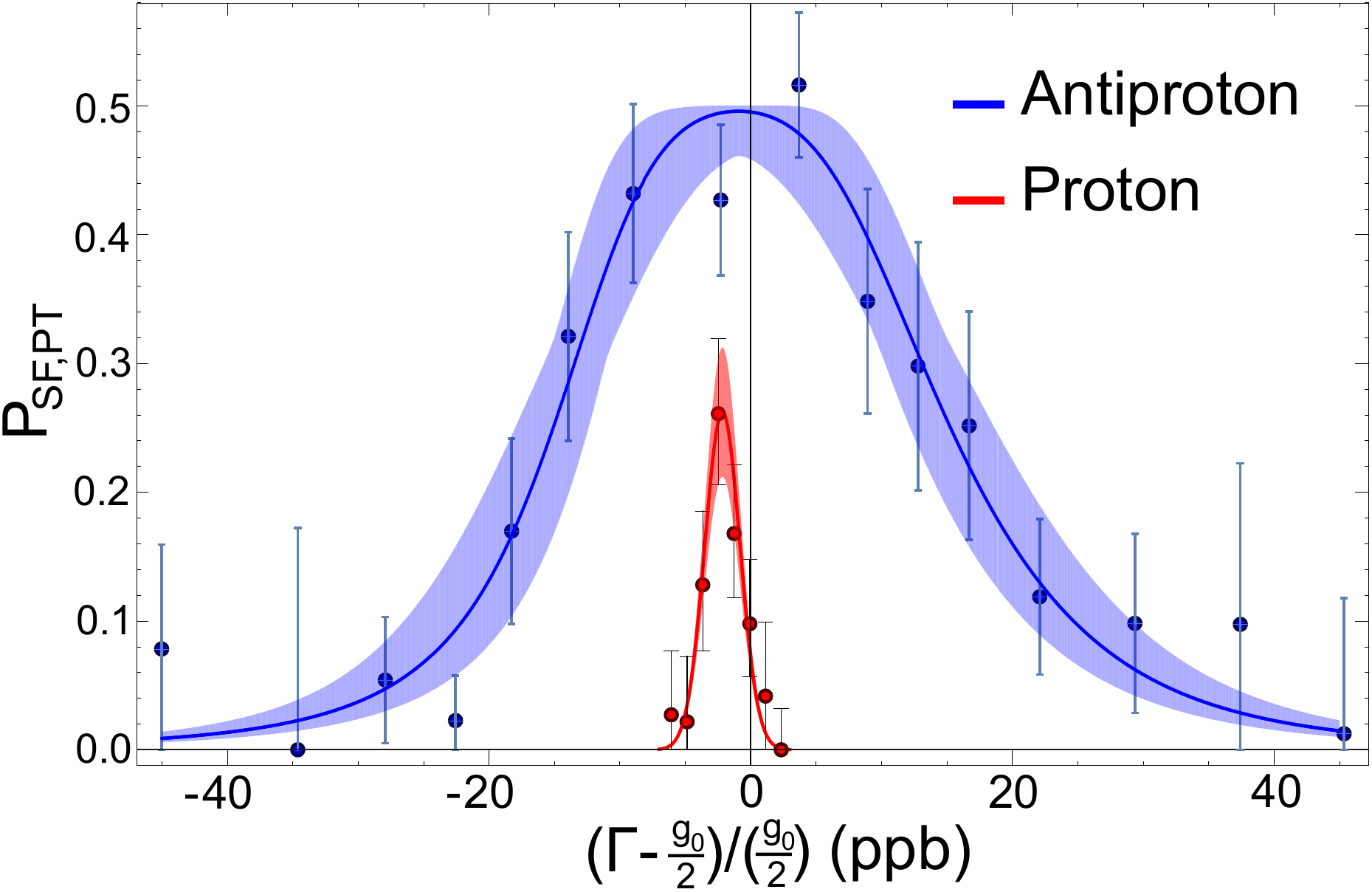}
\caption{\textbf{Spin-flip resonances of the antiproton and proton magnetic moment measurements \cite{SmorraNature,SchneiderScience2017}.} The lineshape function $P_{\mathrm{SF,PT}}(\Gamma$) of the spin-flip resonance in the precision trap as function of the tested frequency ratio $\Gamma$ is shown for protons and antiprotons in red and blue, respectively. The solid line corresponds to the parameters with the maximum likehood, and the shaded area corresponds to the largest change when the lineshape parameters are changed within 1 standard deviation. The antiproton resonance uses an improved lineshape function compared to reference \cite{SmorraNature}. The measurement is referenced to the BASE proton magnetic moment measurement in 2014: $g_0/2 = 2.792847350(9)$ \cite{MooserNature}. }
\label{fig:Resonances}
\end{figure}

The Penning trap system of the BASE collaboration which was used to perform the 0.3 ppb proton magnetic moment measurement is shown in Fig.~\ref{fig:ProtonTrap} \cite{SchneiderScience2017}. 
A single proton, loaded in-trap with a cryogenic field emission point, is prepared for spin-transition spectroscopy by cooling of the cyclotron mode in the precision trap with a resonant cyclotron detector with an effective temperature of 3.2$\,$K. 
We select a low-energy thermal state ($E_+/k_B < 0.6\,$K) by detuning the image-current detector, and transporting the proton adiabatically to the analysis trap, where the cyclotron energy is determined by measuring the axial frequency shift, see Eq.~(\ref{eq:CSGE}).
The cooling cycle is repeated until the cyclotron energy is sufficiently low to observe single spin transitions.
We identify the initial spin state of the cold proton by driving resonantly spin transition and measuring the resulting axial frequency shift \cite{MooserPRL2013}.
Subsequently, the proton is transported back into the precision trap, where we measure the cyclotron frequency $\nu_c$ with the sideband method described in \cite{CornellSB} and drive a spin transition at the frequency ratio $\Gamma = \nu_{\mathrm{rf}}/\nu_c$. 
Subsequently, we cool the cyclotron mode of the proton, since the sideband measurement couples the cyclotron mode to an axial image-current detector with an effective temperature of $T_+ \sim 350\,$K. 
Once the proton is cooled below the energy threshold for single spin-flip detection, we shuttle the proton back to the analysis trap and determine the final spin state.
This enables determining whether the spin in the homogeneous magnetic field of the precision trap was flipped.  
This spin flip determination also initializes the next measurement cycle.
For the latest measurement, about 1300 of these measurement cycles were performed and the proton $g$-factor was extracted from a maximum-likelihood analysis by matching the line-shape function \cite{BrownGeoniumLineshape} to the measured spin-flip probability $P_{\mathrm{SF,PT}}(\Gamma)$, as shown in Fig.~\ref{fig:Resonances}. 

The most significant uncertainty of this measurement is due to measurement statistics (268$\,$ppt) resulting from the slow cycle time due to the inefficient cooling procedure for the cyclotron mode. 
The largest systematic limits resulted from the image-charge shift, i.e. the back-interaction of the image-charge in the trap electrodes with the proton itself \cite{ImageChargeShift,ImageChargeShift2}, which causes a shift of the measured cyclotron frequency with about -98(3) ppt. An additional uncertainty of 80 ppt was added to account for potential shifts from the fitting procedure of the axial frequency.
Including all systematic effects, the result of this proton magnetic moment measurement is
\begin{eqnarray}
\mu_p/\mu_N =2.792\,847\,344\,62(83),
\end{eqnarray} 
which has a relative uncertainty (1 s.d.) of $\sim 300\,$ppt.
Further details on the experiment, the proton $g$-factor measurement and the data analysis are presented in reference \cite{SchneiderScience2017}.

\section{Precision measurement of the antiproton magnetic moment}
\begin{figure}[h!]
\centering
\includegraphics[width= 0.984 \textwidth]{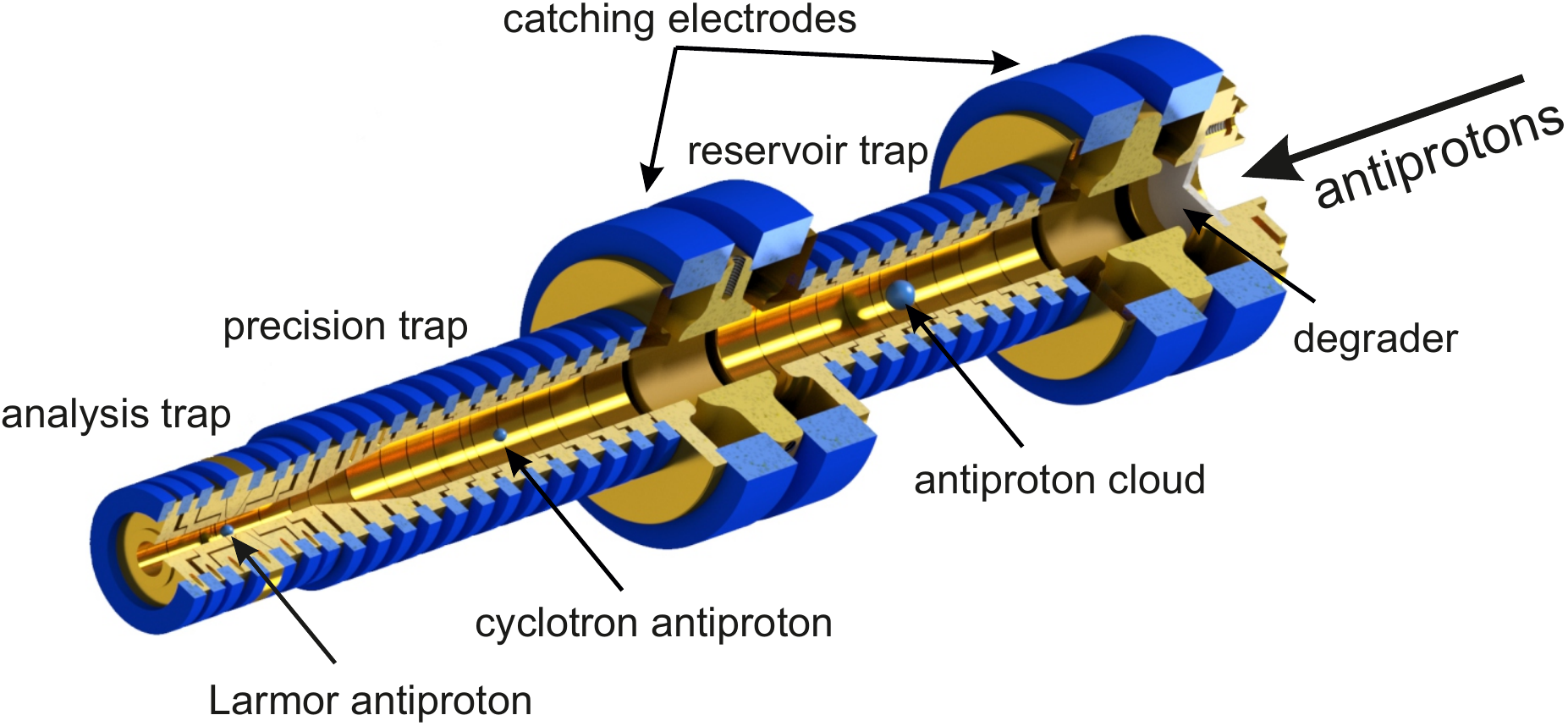}
\caption{\textbf{Trap system for the antiproton magnetic moment measurement}. The trap electrodes and the sapphire rings forming the Penning traps are shown in gold and blue, respectively. A cloud of antiprotons in trapped in the reservoir trap, which supplies the other traps with single antiprotons for the precision measurements. It is surrounded by the catching electrodes, which confine antiprotons after they passed through the degrader by the application of high voltage pulses after the initial injection from the AD. The cold "Larmor antiproton" is shown in the anaysis trap, which has a ferromagnetic ring electrode to identify the spin state, and the second "cyclotron antiproton" is shown in the center of the precision trap, where it is used for cyclotron frequency measurements. More details about the trap system are found in reference \cite{SmorraEPJST2015}.}
\label{fig:PBarTrap}
\end{figure}

The antiproton $g$-factor measurement \cite{SmorraNature} was conducted in the BASE Penning-trap system at CERN \cite{SmorraEPJST2015}, which is shown in Fig.~\ref{fig:PBarTrap}. 
It consists of four Penning traps, two of them are the analysis trap and the precision trap for the double-trap measurement. 
In addition, there is an antiproton reservoir trap and a cooling trap (not shown in Fig.~\ref{fig:PBarTrap}) to accelerate the cooling of the cyclotron mode \cite{SmorraEPJST2015}.
The reservoir trap serves as a single antiproton source for the precision measurements \cite{BASETDR}.
A cloud of antiprotons initially prepared from a single antiproton decelerator pulse is stored in this trap and single antiprotons are extracted non-destructively by splitting the potential well with voltage ramps \cite{SmorraIJMS2015}. 
This technique was developed by the BASE collaboration, and was one of the key methods for the fast progress of the antiproton measurements.
The BASE collaboration managed to run single particle measurements in the trap system continuously for 405 days without reloading - even during shutdown periods of the antiproton decelerator \cite{Sellner2017}. \\

The antiproton magnetic moment measurement faced additional challenges compared to the one of the proton due to the higher radiofrequency noise level in the AD.
The noise increases the cyclotron transition rate in the analysis trap, so that the cyclotron energy limit to reach a sufficient axial frequency stability for the single spin-state detection is lower, $E_+/k_B < 0.2\,$K \cite{BorchertPRL}. Further, the effective temperature of the cyclotron detector of 12.8$\,$K for cyclotron cooling was also limited by radiofrequency noise. Both effects increased the cyclotron cooling time to about 12 hours per cooling cycle.
To circumvent a significant decrease in the measurement statistics, the BASE collaboration invented a new two-particle measurement method, where the first particle is cooled to $E_+/k_B < 50\,$mK and is exclusively used for the spin-transition spectroscopy, whereas the second particle is only used for the cyclotron frequency measurement in the precision trap \cite{SmorraNature}.
This novel scheme has been the essential step to realize the antiproton magnetic moment measurement, since it prevents  the cyclotron mode of the first particle from being heated during the measurement cycle.
In this measurement, cyclotron energy changes of the first particle occur only due to the residual interaction with the detuned cyclotron detector while the particle is in the precision trap. 
This results in a random-walk of the cyclotron energy with a standard deviation of $\Xi_{\Delta E_+}/k_B \sim 22\,\sqrt{N}\,$mK, where $N$ is the number of measurement cycles.
In this way, 75 measurement cycles were performed on average before cyclotron mode cooling became necessary.
In total, 933 data points were recorded, which resulted in the blue resonance shown in Fig.~\ref{fig:Resonances} and extracted the antiproton $g$-factor with a statistical uncertainty of $\sim 1.1\,$ppb.
The two-particle method boosts the measurement statistics by more than a factor of 5, but as a trade-off one has to consider systematic uncertainties due to the different orbits of the two particles in the precision trap.
The most significant shifts scale with the residual linear and quadratic magnetic field gradients and impose in total a systematic shift of 0.44$\,$ppb and a systematic uncertainty of $\sim 1.0\,$ppb.
In total, considering all systematic effects in reference \cite{SmorraNature} and from using an improved lineshape model which has recently been developed (publication in progress), the resulting antiproton magnetic moment is:
\begin{eqnarray}
\mu_{\overline{p}}/\mu_N = -2.7928473443(46),
\end{eqnarray}
which differs by less than 5$\%\,$ of the uncertainty from the result in reference \cite{SmorraNature}, and has a slightly increased relative uncertainty of 1.7$\,$ppb for the 68$\%$ C.L., and 2.9$\,$ppb~for the 95$\%$ C.L.


\section{Interpretation}
The recent proton/antiproton magnetic moment measurements have improved the uncertainty of the proton magnetic moment by a factor of 33 compared to the hydrogen maser measurement \cite{Winkler1972}, and the antiproton magnetic moment by more than a factor 10$^6$ compared to the spectroscopy of antiprotonic helium spectroscopy \cite{Pask2009}, and more than a factor 3000 compared to the best other competing Penning trap measurement \cite{JackAntiproton}.
By combining the two latest magnetic moment measurements, we obtain the following limit on the time-average difference of the proton and antiproton magnetic moments:
\begin{eqnarray}
\frac{\delta \mu}{\mu_N} = \left(\frac{\mu_p}{\mu_N}+\frac{\mu_{\overline{p}}}{\mu_N}\right) = 0.3 (8.3) \times 10^{-9},
\end{eqnarray}
where the uncertainty in brackets is for 95$\,\%$ C.L. 
Our result shows no significant deviation from the zero value predicted by CPT invariance and support CPT symmetry within the resolution of our measurements. 
Consequently, we can set limits on all CPT-odd interactions which produce a time-average difference of the Larmor frequency or the magnetic moments, such as the non-minimal Standard Model Extension (SME) \cite{Ding2016} and a certain CPT-odd dimension-five operator \cite{Stadnik2014}. We obtain the following constraints (95$\,\%$ C.L.) on the combination of six non-minimal SME coefficients from the uncertainty of $\delta g$ of our two latest measurements:
\begin{eqnarray}
\left| \tilde{b}^{Z}_{p} \right| < 8.1 \times 10^{-25}\, \mathrm{GeV}, \\
\left| \tilde{b}^{XX}_{F,p}+\tilde{b}^{YY}_{F,p} \right| < 4.6 \times 10^{-9}\, \mathrm{GeV}^{-1}, \\
\left| \tilde{b}^{ZZ}_{F,p} \right| < 3.3 \times 10^{-9}\, \mathrm{GeV}^{-1}, \\
\left| \tilde{b}^{*Z}_{p} \right| < 1.5 \times 10^{-24}\, \mathrm{GeV}, \\
\left| \tilde{b}^{*XX}_{F,p}+\tilde{b}^{*YY}_{F,p} \right| < 3.1 \times 10^{-9}\, \mathrm{GeV}^{-1}, \\
\left| \tilde{b}^{*ZZ}_{F,p} \right| < 1.1 \times 10^{-8}\, \mathrm{GeV}^{-1}. 
\end{eqnarray}
The last three combinations of coefficients represent the most stringent constraints for these  antiproton coefficients \cite{CPTTables}. In addition, we set the following limit (95$\,\%$ C.L.) on a possible magnetic moment splitting for protons and antiprotons $f_p^0$ caused by a certain dimension-five operator described in reference \cite{Stadnik2014}:
\begin{eqnarray}
f_p^0 = \frac{\delta g}{2} \frac{\mu_N}{2} < 4.5 \times 10^{-12} \mu_B.
\end{eqnarray}

Despite the factor 3000 improvement in the sensitivity of testing CPT invariance in the baryon sector, there is still no hint how the matter-antimatter asymmetry was created, neither in the measurements of the BASE collaboration nor in the results of many other high-precision tests of the fundamental interactions \cite{NPAM-Review}. Currently, the BASE collaboration is targeting to increase the measurement precision by a factor 10 to 100 in the next years. The ongoing technical developments to reach this improved sensitivity are described in the sections 8 and 9. These technical advances will allow revisiting tests of the CPT invariance in the baryon sector with one or two orders improved sensitivity.


\section{High-precision comparisons of the proton-to-antiproton charge-to-mass ratio}
A second particle property which can be measured with high precision in Penning traps is the charge-to-mass ratio. To this end, one measures the cyclotron-frequency ratio of two particles in the same magnetic field, which results in a direct comparison of their charge-to-mass ratios. In our case, we measure the antiproton-to-proton cyclotron frequency ratio:
\begin{equation}
\frac{\nu_{c,\overline{p}}}{\nu_{c,p}} = \frac{(q/m)_{\overline{p}}}{(q/m)_p}.
\end{equation}
Consequently, this measurement provides a second stringent CPT invariance test in the baryon sector. Such measurements were already performed by the TRAP collaboration \cite{JerryReview2006}, with the latest result reported in 1999 having 90$\,$ppt relative uncertainty \cite{JerryAntiproton}. One key technique is to use the negative hydrogen ion as proxy for the proton, since the conversion factor $R=\nu_{c,p}/\nu_{c,H^-}$ is known with 0.2 ppt uncertainty \cite{UlmerNature2015,Higuchi} and since this greatly reduces systematic uncertainties related to inverting the trap voltage. \\

The BASE Penning trap system has also been utilized to conduct such a high-precision comparison of the charge-to-mass ratios of protons and antiprotons \cite{UlmerNature2015}. Compared to these earlier measurements, the cyclotron frequency was measured at lower cyclotron energy in thermal equilibrium with the axial detection system resulting in an effective temperature of about 350$\,$K. 
Further, the BASE collaboration first realized in this measurement a scheme using the fast adiabatic transport to exchange the particles in the measurement trap. 
This method enabled measuring the cyclotron-frequency ratio with a cycle time of 4 minutes, about a factor 60 faster than in earlier measurements. 
The cycle time was limited by the necessity to average the magnetic field fluctuations over one deceleration cycle of the antiproton decelerator. 
This fast and entirely non-destructive measurement scheme is meanwhile also used for other precision charge-to-mass ratio measurements \cite{HeisseProton}.  \\
    
In 2014, the BASE collaboration recorded about 6500 cyclotron frequency ratios with this measurement scheme in 35 days, which resulted in a statistical uncertainty of 62 ppt. The major systematic uncertainty resulted from the interplay of the residual magnetic field gradient in the measurement trap with a tiny change of $\sim 30\,$nm of the equilibrium position of the trap caused by a trap voltage adjustment for the negative hydrogen ion. This resulted in a correction of 119(20)$\,$ppt. The final result of the charge-to-mass ratio comparison was:
\begin{eqnarray}
\frac{(q/m)_{\overline{p}}}{(q/m)_p}+1 = 1(69) \times 10^{-12},
\end{eqnarray}
which is also consistent with CPT invariance. \\

The charge-to-mass ratio comparison is sensitive to gravitational anomalies of antiprotons, since any difference in the gravitational binding energy would result in a different gravitational redshift of the proton and antiproton cyclotron frequencies \cite{HughesHolzscheiter}. 
The cyclotron-frequency ratio including the gravitational redshift by the gravitational potentials $U$ and $\alpha_g U$ for protons and antiprotons, respectively, is given as:
\begin{eqnarray}
\frac{\nu_{c,\overline{p}}}{\nu_{c,p}} = \frac{(q/m)_{\overline{p}}}{(q/m)_p} (1+ 3(\alpha_g-1)U c^{-2}).
\end{eqnarray}
Here, $\alpha_g$ is the parameter characterizing the modified gravitational acceleration of antiprotons, $m_p$ the proton mass, and $m_{\overline{p}}$ the antiproton mass.
Limits derived from such comparisons are however ambiguous, since they are simultaneously a test of CPT invariance and gravitational anomalies. Consequently, an intrinsic charge-to-mass ratio difference from breaking the CPT invariance cannot be disentangled from an anomalous gravitation. 
 Assuming CPT invariance to be valid and using for $U$ the gravitational potential of the local galactic supercluster, which is an approach consistent with earlier literature \cite{HughesHolzscheiter}, but which has been controversially discussed \cite{NietoGoldman}, limits on the gravitational anomaly parameter for antiprotons were extracted from this measurement: $(\alpha_g - 1) < 8.7\cdot 10^{-7}$ \cite{UlmerNature2015}.\\

\section{Improved antiproton lifetime limits}
Another fundamental property of antiprotons is their lifetime $\tau_{\overline{p}}$, which is required by CPT invariance to be identical to the proton lifetime $\tau_p$. 
Searching for proton decays is of high interest in the scope of finding a baryon-number violating process, therefore searches based on huge samples with more than 1000 tons of water were made looking for a disappearance signal. 
These experiment found $\tau_p > 10^{29}\,$y \cite{ProtonDisappearance}, and even higher constraints have been obtained for specific decay channels \cite{ProtonAppearance}. On the contrary, due to the difficulty in producing and storing antiprotons, $\tau_{\overline{p}} > 0.3\,$y was until recently the best limit for an antiproton disappearance signal \cite{JerryLifeTime}. \\

The trap system of the BASE collaboration enables to search for a disappearance signal by storing clouds of antiprotons in the reservoir trap, and counting the amount of confined particles over time by making use of the lineshape of the image-current signal \cite{Sellner2017}. 
Based on experiments from July 2014 until December 2016, the BASE collaboration has not observed any antiproton loss due to annilihation with residual gas. 
This enabled our collaboration to set a limit of $\tau_{\overline{p}}>10.2\,$yrs (68$\,\%$ C.L.) on the total lifetime of antiprotons, which represents the best laboratory measurement so far \cite{Sellner2017}. 
Other laboratory constraints were set by the APEX collaboration, which was operating at the Fermilab antiproton storage ring. APEX searched for antiproton decays in the storage ring producing negative leptons as decay product, and obtained limits for 13 decay channels in the range of 200$\,$yrs to $7 \cdot10^5\,$yrs \cite{APEX1}. 
In contrast, the measurement of our collaboration constrains all possible decay channels, including also decays into negative kaons \cite{KaonDecayChannel1,KaonDecayChannel2} and other ``dark decay channels'', which were invisible to the APEX detector. 
However, astrophysical bounds on the antiproton lifetime suggest that $\tau_{\overline{p}} \gtrsim 10^6\,$yrs \cite{APEX2}, so that improved experimental techniques are required to exceed these bounds in laboratory experiments. 
The existing methods may further improve the present experimental limit by a factor 100 by increasing the number of antiprotons in the reservoir trap. 
Further increases require new approaches for efficient non-destructive counting of a large cloud of trapped particles. 
In case annihilation events are observed at a higher lifetime and vacuum sensitivity, we also require an independent vacuum measurement, which could be based on the recombination rate of highly-charged ions to constrain the annihilation rate with residual gas \cite{Sellner2017}. \\

\section{Coupling trap}
One of the major limitations in proton/antiproton $g$-factor measurements is the impact of the cyclotron cooling time on the measurement statistics. 
Recooling protons and antiprotons below the energy threshold for single spin-flip detection requires on average 2 hours and 12 hours, respectively, in the latest measurements.
The resistive cooling technique is limited by the detector temperature to a few Kelvin, whereas laser cooling of trapped ions has already reached the motional ground state of cyclotron mode in a Penning trap \cite{Thompson2016}.
Atomic ions without suitable laser-cooling transition can be sympathetically cooled by co-trapped ions, however we require also the cooling of a single antiproton, which cannot be performed with co-trapped ions. \\

\begin{figure}[h!]
\centering
\includegraphics[width= 0.5 \textwidth]{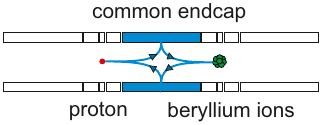}
\caption{\textbf{Image-current interaction in a common-endcap two-trap system}. The proton (red) and a cloud of beryllium ions (green) interact by image-currents though a common-endcap electrode (blue). The blue lines visualize the Coulomb interaction with the image currents in the common-endcap electrode.}
\label{fig:CETrap}
\end{figure}

To reduce the cooling limits for single protons and antiprotons, the BASE collaboration follows an early idea by Heinzen and Wineland \cite{Wineland1990} and developed a "common-endcap" two-trap system, where trapped particles in separate potential wells interact through image currents induced onto a shared trap electrode. The principle is illustrated in Fig.~\ref{fig:CETrap}.
By matching the axial frequencies of the trapped particles, the image-current interaction leads to a periodic exchange of the energy for both particles in the shared axial mode.
The coupling trap system consists of two orthogonal compensated Penning traps \cite{Jerry5poleTrap} where the electrodes in between the two ring electrodes transmit the image-current signal to couple protons to laser-cooled beryllium ions \cite{Bohman2018}.
The trap system is optimized to have a low capacitance on the common endcap electrode (5 pF) and a short particle-electrode distance (4 mm) to obtain a short energy exchange time. 
We expect to reach an exchange time of 55 s by coupling a cloud of 100 beryllium ions to a single proton \cite{Bohman2018}.
If this goal can be reached, (anti)protons will be deterministically cooled to a cyclotron energy of a few 10 $\,$mK with a significantly reduced cooling time. 
The lower temperature also ensures that the spin-flip detection fidelity is close to 100$\,\%$, and that frequency measurements can be performed at lower temperatures and lower amplitude-dependent frequency shifts.
Consequently, the sympathetic cooling in the common-endcap trap system will be one of the essential improvements to improve the precision of proton/antiproton magnetic moment measurements.

\section{A Transportable Trap for Antiprotons}
Another important limitation in the antiproton measurements is the magnetic field and radiofrequency noise imposed by the operations in the antiproton decelerator facility.
The operation of the antiproton decelerator causes about 1$\,\mu$T peak-to-peak fluctuations in the BASE experimental area, which would, if not suppressed, cause relative shifts of the cyclotron frequency of order 5$\times 10^{-7}$. 
Presently, the BASE collaboration makes use of the magnetic shielding provided by self-shielding superconducting solenoids \cite{Higuchi}, which were recently improved \cite{JackElise2019} to suppress cyclotron-frequency ratio fluctuations to below 2$\,$ppb. 
However, on the long-term, we target to perform antiproton precision measurements in a low-noise environment, such that we can reach an even higher magnetic field stability.
To this end, we target to develop a transportable trap for antiprotons, which can be loaded at the antiproton decelerator/ELENA facility, and can be transported to a high-precision laboratory to conduct single-antiproton measurements with a transportable antiproton reservoir. \\

Compared to earlier attempts \cite{Tseng1993,Wada1993,LewisMedical} and presently active projects \cite{PUMA}, we target to develop a transportable trap for single particle experiments, which requiring only a low amount of antiprotons. 
The transportable trap will form a single antiproton source for single-particle precision measurements based on the reservoir trap technique \cite{SmorraIJMS2015}. We have demonstrated the long-term storage of antiprotons up to more than a year \cite{Sellner2017}, which is sufficient to conduct precision measurements in an offline laboratory. \\

The possibility to make antiproton charge-to-mass ratio and magnetic moment measurements in a high-precision laboratory will lead to further improvements in precision beyond limits imposed by the noisy environment at the production site. 
Further, the possibility to increase the number of antiproton precision experiments enables new kinds of experiments based on simultaneous antiproton cyclotron frequency measurements. This will allow testing Lorentz-invariance within the framework of the Standard Model Extension \cite{Ding2016}, and enable searches for topological dark-matter with a network of antimatter clocks \cite{Derevianko}, such as presently performed with magnetometers \cite{NuclSpinComag} and atomic clocks \cite{NPAM-Review}. Another possibility is to search for changes in the gravitational redshift of the antiproton cyclotron frequency by an active modification of the gravitational potential by measuring at the cyclotron-frequency ratio of antiprotons and negative hydrogen ions at different altitudes. However, this requires a further advance in measurement precision by a factor of $\sim$ 70 to reach the sensitivity to probe antiproton gravity at a height difference of 300$\,$m. \\

\section{Magnetic Moment Measurements of He-3 ions}
Based upon the developments for the proton and the antiproton magnetic moment measurements we started the construction of a new experiment, which aims at the first direct high-precision measurement of the nuclear magnetic moment of $^3$He$^{2+}$ and a high-precision measurement of the ground-state hyper-fine splitting in a magnetic field of $^3$He$^{+}$. 
To this end sympathetic laser-cooling, a novel analysis-trap as well as a novel schema for the detection of the spin-state will be applied for the first time. \\

The measurement of the nuclear magnetic moment of $^3$He$^{2+}$ will establish hyperpolarized $^3$He as an independent magnetometer, which compared to H$_2$O probes has smaller systematic uncertainties concerning sample shape, impurities, environmental dependencies and exhibits no line-shape distortion with a poorly defined center \cite{Reid,Grange,Fei,Chupp}. Furthermore, the diamagnetic shielding correction of $^3$He is the one which is most precisely known for all atoms \cite{Rudzinski}. 
Making use of optical pumping and long relaxation times in low-pressure $^3$He gas cells, relative precisions of 10$^{-12}$ within seconds were already achieved \cite{Nikiel}. 
However, limited by theoretical diamagnetic shielding corrections to 12 ppb, the $^3$He probes do not provide an absolute calibration independent of H$_2$O, because the magnetic moment of $^3$He was only determined by measuring the $^3$He NMR frequency relative to the NMR frequency of H$_2$O \cite{Flowers}. 
Thus, so far, any measurement of a given magnetic field using $^3$He and H$_2$O cannot be considered as independent and uncorrelated but rather represents a revaluation of existing NMR frequency-ratio measurements used to determine the $^3$He$^{2+}$ magnetic moment. 
In addition, deviations in NMR frequency comparisons compared to accepted values are reported \cite{Jackowski}, which can be explained by inconsistencies in proton shielding corrections or a shift of the nuclear magnetic moment of $^3$He at the 100 ppb level. 
The planed measurement will thus eventually overcome the limitation of a calibration correlated to H$_2$O NMR probes, resolve deviations in reported nuclear magnetic measurements data and allow for a ppb test of theoretical diamagnetic shielding corrections.\\

The measurement of the ground-state hyper-fine splitting in a magnetic field of $^3$He$^{+}$ will provide high-precision determinations of the bound electronic and nuclear magnetic moments in a light hydrogen like atom. 
In addition, it will give access to the zero-field ground-state hyper-fine splitting, which strongly depends on nuclear structure effects, e.g. nuclear polarizability and the Zeemach radius. 
Currently, the most precise measurement results from spectroscopy in a radio-frequency quadrupole ion trap \cite{dehmelt}. 
The measurement was limited to relative precision of 1 ppb by the second-order Doppler effect in a room temperature apparatus. 
We aim to improve the precision by making use of single ion spectroscopy in a cryogenic Penning-trap system.




\section{Conclusion}
We have presented the recent improvements of the BASE collaboration in comparing the fundamental properties of protons and antiprotons. The measurements presented here support the CPT invariance as fundamental symmetry of the Standard Model up to the present measurement resolution. New methods are presently being developed to make further improvements in the sensitivity of testing CPT invariance, and to conduct a measurement of the He-3 magnetic moment.

\ack
We acknowledge all former and present members of the BASE collaboration for their contributions to the results presented here. 
In particular we would like to thank Klaus Blaum and Stefan Ulmer for their continuous support and their comments to this manuscript. 
We acknowledge financial support by RIKEN, the RIKEN Foreign Postdoctoral Researcher program, RIKEN Incentive Research Project Funding, the Max-Planck Society, the Max-Planck-RIKEN-PTB Center for Time, Constants and Fundamental Symmetries, and the CERN fellowship program.  We acknowledge support from CERN, in particular, from the AD operation team and the CERN cryolab team. \\

\appendix


\section*{References}
\begin{thebibliography}{9}

\bibitem{Weinberg2018}
Weinberg S 
2018 
Essay: Half a Century of the Standard Model 
\textit{Phys.~Rev.~Lett.~}\textbf{121} 
220001%

\bibitem{WuExperiment}
Wu C S, Ambler E, Hayward R W, Hoppes D D and Hudson R P 
1957 
Experimental Test of Parity Conservation in Beta Decay 
\textit{Phys.~Rev.~}\textbf{105} 1413-5

\bibitem{CroninKaon}
Christenson J H, Fitch V L, Cronin J W and Turlay R 
1964
Evidence for 2$\pi$ decay of K$_2^\circ$ meson
\textit{Phys.~Rev.~Lett.~}\textbf{13} 138-40

\bibitem{CPT1}
L\"uders G 
1957 
Proof of the TCP theorem
\textit{Ann.~Phys.~}\textbf{2} 1-15

\bibitem{BaryonAsymmetry}
Cohen A G, De Rujula A and Glashow S L
1998
A Matter-Antimatter Universe?
\textit{Astroph.~J.~}\textbf{495}
539-49

\bibitem{BAU-review}
Dine M and Kusenko A 
2004
Origin of the matter-antimatter asymmetry
\textit{Rev.~Mod.~Phys.~}\textbf{76}, 1-30

\bibitem{CPTTables}
Kostelecky V A and Russell N 
2019
Data table for Lorentz and CPT violation
\textit{arXiv} 0801.0287v12 [hep-ph]


\bibitem{Maury1999HypInt} 
Hemery J Y and Maury S
1999
The Antiproton Decelerator: Overview
\textit{Nucl.~Phys.~}A
\textbf{655} 
345-52

\bibitem{JerryAntiproton} 
Gabrielse G, Khabbaz A, Hall D, Heimann C, Kalinowsky H and Jhe W
1999
Precision mass spectroscopy of the antiproton and proton using simultaneously trapped particles
\emph{Phys.~Rev.~Lett.~}\textbf{82} 
3198-201

\bibitem{JackAntiproton} 
DiSciacca J \textit{et al.} 
2013 
One-Particle Measurement of the Antiproton Magnetic Moment
\textit{Phys.~Rev.~Lett.~}\textbf{110} 130801

\bibitem{UlmerNature2015} 
Ulmer S \textit{et al.}
2015
High-precision comparison of the antiproton-to-proton charge-to-mass ratio
\textit{Nature} 
\textbf{524} 
196-9

\bibitem{SmorraNature} 
Smorra, C.~\emph{et al}.
2017
A parts-per-billion measurement of the antiproton magnetic moment
\textit{Nature} 
\textbf{550} 
371-4

\bibitem{ALPHA} 
Ahmadi M \emph{et al.}
2018
Chacterization of the 1S-2S transition in antihydrogen
\textit{Nature}
\textbf{557} 
71-4

\bibitem{ALPHA2}
Ahmadi M \textit{et al.}
2017
Observation of the hyperfine spectrum of antihydrogen
\textit{Nature}
\textbf{548}
66-9

\bibitem{ASACUSAHBarBeam}
Kuroda N \textit{et al.}
2014
A source of antihydrogen for in-flight hyperfine spectroscopy
\textit{Nat.~Commun.~}\textbf{5}
3089

\bibitem{ASACUSAHydrogen}
Malbrunot C \textit{et al.}
2018
The ASACUSA antihydrogen and hydrogen program: results and prospects
\textit{Phil.~Trans.~Roy.~Soc.~}\textbf{376}
20170273


\bibitem{Masaki} 
Hori M, Aghai-Khozani H, Soter A, Barna D, Dax A, Hayano R, Kobayashi T, Murakami Y, Todoroki K, Yamada H, Horvath D and Venturelli L
2016
Buffer-gas cooling of antiprotonic helium to $1.5$ to $1.7\,$K, and antiproton-to-electron mass ratio 
\textit{Science} 
\textbf{354}
610-4

\bibitem{WalzHbarPlus}
Walz J, Pittner H, Herrmann M, Fendel P, Henrich B and H\"ansch T W
2003
Cold antihydrogen atoms
\textit{Appl.~Phys.~}B
\textbf{77}
713-7


\bibitem{GBAR}
Perez P \textit{et al.}
2015
The GBAR antimatter gravity experiment
\textit{Hyp.~Int.~}\textbf{233}
21-7

\bibitem{DehmeltHbar2Ion}
Dehmelt H
1995
Economic synthesis and precision spectroscopy of anti-molecular hydrogen ions in Paul trap
\textit{Phys.~Scr.~}\textbf{T59} 
423

\bibitem{MyersHbar2Ion}
Myers E G
2018
CPT tests with the antihydrogen molecular ion
\textit{Phys.~Rev.~}A
\textbf{98}
010101

\bibitem{AEgIS}
Aghion S \textit{et al.}
2014
A moire deflectometer for antimatter
\textit{Nat.~Commun.~}\textbf{5}
4538

\bibitem{ALPHAg}
Bertsche W A
2018
Prospects for comparison of matter and antimatter gravitation with ALPHA-g
\textit{Phil.~Trans.~Roy.~Soc.~}\textbf{376}
20170265

\bibitem{ALPHAGrav}
Charman A E \textit{et al.}
2013
Description and first application of a new technique to measure the gravitational mass of antihydrogen
\textit{Nat.~Commun.~}\textbf{4}
1785

\bibitem{ELENA}
Bartmann W \textit{et al.}
2018
The ELENA facility
\textit{Phil.~Trans.~Roy.~Soc.~}\textbf{376}
20170266


\bibitem{Winkler1972}
Winkler P F, Kleppner D, Myint T and Walther F G 
1972
Magnetic moment of the proton in Bohr magnetons 
\textit{Phys.~Rev.~}A \textbf{5}, 83-114

\bibitem{Pask2009}
Pask T \textit{et al}. 
2009 
Antiproton magnetic moment determined from the HFS of (p)over-barHe(+)
\textit{Phys.~Lett.~}B \textbf{678} 55-9

\bibitem{DehmeltCSG} 
Dehmelt H and Ekstr\"om P 
1973
Proposed $g$-2 delta-omegaz experiment on single stored electron or positron
\textit{Bull.~Am.~Phys.~Soc.~}\textbf{18} 727%

\bibitem{Monoelectron}
Wineland D, Ekstr\"om P and Dehmelt H 
1973
Monoelectron Oscillator
\textit{Phys.~Rev.~Lett.~}\textbf{31} 1279-82

\bibitem{Dehmeltg-2}
Van Dyck R S, Schwinberg P B and Dehmelt H G 
1987
New High-Precision Comparison of Electron and Positron $g$ Factors
\textit{Phys.~Rev.~Lett.~}\textbf{59} 26-9

\bibitem{DehmeltCPT}
Dehmelt H, Mittleman R, Van Dyck R S and Schwinberg P 
1999
Past electron-positron $g$-2 experiments yielded sharpest bound on CPT violation for point particles
\textit{Phys.~Rev.~Lett.~}\textbf{83} 4694-6

\bibitem{Quint2004}
Quint W, Alonso J, Djekic S, Kluge H J, Stahl S, Valenzuela T, Verdu J, Vogel M and Werth G 
2004
Continuous Stern-Gerlach effect and the magnetic moment of the antiproton
\textit{Nucl.~Instr.~Meth.~}B \textbf{214} 207-10

\bibitem{haeffner2003double} 
H\"affner H, Beier T, Djekic S, Hermanspahn N, Kluge H J, Quint W, Stahl S, Verdu J, Valenzuela T and Werth G 
2003
Double Penning trap technique for precise $g$ factor determinations in highly charged ions
\textit{Eur.~Phys.~J.~}D \textbf{22} 163-82

\bibitem{UlmerPRL2011} 
Ulmer S, Rodegheri C C, Blaum K, Kracke H, Mooser A, Quint W, and Walz J 
2011
Observation of Spin Flips with a Single Trapped Proton
\textit{Phys.~Rev.~Lett.~}\textbf{106} 253001

\bibitem{MooserPRL2013} 
Mooser A, Kracke H, Blaum K, Br\"auninger S A, Franke K, Leiteritz C, Quint W, Rodegheri C C, Ulmer S, and Walz J 
2013
Resolution of Single Spin Flips of a Single Proton
\textit{Phys.~Rev.~Lett.~}\textbf{110} 140405

\bibitem{SmorraPLB2017} 
Smorra C \textit{et al.} 
2017 
Observation of individual spin quantum transitions of a single antiproton
\textit{Phys.~Lett.~}B \textbf{769} 1-6 

\bibitem{CCRodegheri2012} 
Rodegheri C C, Blaum K, Kracke H, Kreim S, Mooser A, Quint W, Ulmer S and Walz J 
2012
An experiment for the direct determination of the $g$-factor of a single proton in a Penning trap
\textit{New J.~Phys.~}\textbf{14} 063011

\bibitem{SmorraEPJST2015} 
Smorra C \emph{et al}. 
2015 
BASE - The Baryon Antibaryon Symmetry Experiment,
\textit{Eur.~Phys.~J.~Special Topics} \textbf{224} 3055-108

\bibitem{JackProton} 
DiSciacca J and Gabrielse G 
2012
Direct Measurement of the Proton Magnetic Moment
\textit{Phys.~Rev.~Lett.~}\textbf{108} 153001

\bibitem{HiroNC2017} 
Nagahama H \textit{et al.}
2017
Sixfold improved single particle measurement of the magnetic moment of the antiproton
\textit{Nat.~Commun.~}\textbf{8} 14084

\bibitem{MooserNature} 
Mooser A, Ulmer S, Blaum K, Franke K, Kracke H, Leiteritz C, Quint W, Rodegheri C C, Smorra C and Walz J
2014
Direct high-precision measurement of the magnetic moment of the proton
\textit{Nature} \textbf{509} 596-9

\bibitem{SchneiderScience2017}
Schneider G \emph{et al.}
2017
Double-trap measurement of the proton magnetic moment at 0.3 parts per billion precision
\textit{Science} 
\textbf{358} 
1081-4

\bibitem{Ding2016} 
Ding Y and Kostelecky V A 
2016
Lorentz-violating spinor electrodynamics and Penning traps
\textit{Phys.~Rev.~}D
\textbf{94} 
056008

\bibitem{Stadnik2014}
Stadnik Y V, Roberts B M and Flambaum V V 2014 
Tests of CPT and Lorentz symmetry from muon anomalous magnetic dipole moment
\textit{Phys.~Rev.~}D 
\textbf{90} 
045035

\bibitem{Sellner2017}
Sellner S \textit{et al.}
2017
Improved limit on the directly measured antiproton lifetime
\textit{New J.~Phys.~}\textbf{19} 
083023

\bibitem{Wineland}
Heinzen D J and Wineland D J 1990
Quantum-limited Cooling and Detection of Radio-frequency Oscillations by Laser-cooled Ions 
\textit{Phys.~Rev.~}A 
\textbf{42} 
2977-94 

\bibitem{Bohman2018}
Bohman M \emph{et al.}
2018
Sympathetic cooling of protons and antiprotons with a common endcap Penning trap
\textit{J.~Mod.~Opt.~}\textbf{65} 
568-76

\bibitem{QLEDS2018}
Meiners T \emph{et al.}
2018
Towards sympathetic cooling of single (anti-)protons
\textit{Hyp.~Int.~}\textbf{239} 
26%




\bibitem{Hanneke2008} 
Hanneke D, Fogwell S and Gabrielse G
2008 
\textit{Phys.~Rev.~Lett.~}\textbf{100}
120801

\bibitem{SturmAtoms} 
Sturm S, Vogel M, K\"ohler-Langes F, Quint W, Blaum K and Werth G 
2017
High-Precision Measurements of the Bound Electron’s Magnetic Moment
\textit{Atoms} 
\textbf{5} 
4%

\bibitem{HeisseProton}
Hei\ss e F, K\"ohler-Langes F, Rau S, Hou J, Junck S, Kracke A, Mooser A, Quint W, Ulmer S, Werth G, Blaum K and Sturm S 2017
High-Precision Measurement of the Proton’s Atomic Mass
\textit{Phys.~Rev.~Lett.~}\textbf{119} 
033001 

\bibitem{MyersAtoms} 
Myers E G 
2019
High-Precision Atomic Mass Measurements for Fundamental Constants
\textit{Atoms}
\textbf{7} 
37%

\bibitem{Brown} 
Brown L S and Gabrielse G 
1986
Geonium theory: Physics of a single electron or ion in a Penning trap
\textit{Rev.~Mod.~Phys.~}\textbf{58}
233-311

\bibitem{Wine} 
Wineland D J and Dehmelt H G
1975
Principles of the stored ion calorimeter
\textit{J.~Appl.~Phys.~}\textbf{46} 
919-30

\bibitem{UlmerPRL5Dip} 
Ulmer S, Blaum K, Kracke H, Mooser A, Quint W, Rodegheri C C and Walz J
2011
Direct Measurement of the Free Cyclotron Frequency of a Single Particle in a Penning Trap
\textit{Phys.~Rev.~Lett.~}\textbf{107}
103002

\bibitem{HiroRSI} 
Nagahama H \textit{et al.} 
2016
Highly sensitive superconducting circuits at 700 kHz with tunable quality factors for image-current detection of single trapped antiprotons 
\textit{Rev.~Sci.~Instrum.~}\textbf{87} 113305

\bibitem{CornellSB}
Cornell E A, Weisskoff R M, Boyce K R and Pritchard D E
1990
Mode coupling in a Penning trap: π pulses and a classical avoided crossing
\textit{Phys.~Rev.~}A 
\textbf{41} 312-5

\bibitem{Higuchi}
Higuchi T \textit{et al.} 
2018
Progress towards an improved comparison of the proton-to-antiproton charge-to-mass ratios
\textit{Hyp.~Int.~}\textbf{239} 
27%

\bibitem{BorchertPRL}
Borchert M J \textit{et al.}
Measurement of Ultralow Heating Rates of a Single Antiproton in a Cryogenic Penning Trap
\textit{Phys.~Rev.~Lett.~}\textbf{122}
043201


\bibitem{BrownGeoniumLineshape} 
Brown L S
1985
Geonium lineshape
\textit{Ann.~Phys.~}\textbf{159}
62-98


\bibitem{ImageChargeShift}
Porto J V
2001
Series solution for the image charge fields in arbitrary cylindrically symmetric Penning traps
\textit{Phys.~Rev.~}A
\textbf{64}
023403

\bibitem{ImageChargeShift2}
Schuh M, Hei\ss e F, Eronen T, Ketter J, Kohler-Langes F, Rau S, Sega T, Quint W, Sturm S and Blaum K
2019
Image charge shift in high-precision Penning traps
\textit{Phys.~Rev.~}A
\textbf{100}
023411



\bibitem{BASETDR}
Ulmer S, Yamazaki Y, Smorra C, Blaum K, Franke K, Matsuda Y, Nagahama H, Quint W, Walz J, Mooser A and Schneider G
2013
Technical Design Report BASE
\textit{CERN Document Server} 
\textbf{CERN-SPSC-2013-002}
SPSC-TDR-002

\bibitem{SmorraIJMS2015} 
Smorra C \emph{et al.} 
2015
A reservoir trap for antiprotons
\textit{Int.~J.~Mass Spectr.~}\textbf{389}
10-3


\bibitem{NPAM-Review}
Safronova M S, Budker D, DeMille D, Kimball D and Derevianko A 
2018 
Search for new physics with atoms and molecules
\textit{Rev.~Mod.~Phys.~}\textbf{90} 
025008


\bibitem{JerryReview2006} 
Gabrielse G 
2006
Antiproton mass measurements
\textit{Int.~J.~Mass Spectr.~}\textbf{251} 
273-80

\bibitem{HughesHolzscheiter}
Hughes R J and Holzscheiter M H 
1991
Constraints on the gravitational properties of antiprotons and positrons from cyclotron-frequency measurements
\textit{Phys.~Rev.~Lett.~}\textbf{66}
854-7 

\bibitem{NietoGoldman}
Nieto M and Goldman T
1991
The argument against “Antigravity” and the gravitational acceleration of antimatter
\textit{Phys.~Rep.~}\textbf{205} 
221-81 

\bibitem{ProtonDisappearance} 
Ahmed S N \textit{et al.} 
2004
Constraints on Nucleon Decay via Invisible Modes from the Sudbury Neutrino Observatory
\textit{Phys.~Rev.~Lett.~}\textbf{92} 
102004

\bibitem{ProtonAppearance}
Abe K \textit{et al.} 
2017 
Search for proton decay via $p\xrightarrow{}e^+\pi^0$ and $p\xrightarrow{}\mu^+\pi^0$ in 0.31 megaton years exposure of the Super-Kamiokande water Cherenkov detector
\textit{Phys.~Rev.~}D 
\textbf{95} 
012004

\bibitem{JerryLifeTime}
Gabrielse G, Fei X, Orozco L A, Tjoelker R L, Haas J, Kalinowski H, Trainor T A and Kells W 
1990
Thousandfold improvement in the measured antiproton mass
\textit{Phys.~Rev.~Lett.~}\textbf{65} 
1317-20

\bibitem{APEX1}
Geer S H \textit{et al.} 
2000
New Limit on CPT Violation
\textit{Phys.~Rev.~Lett.~}\textbf{84} 
590-3

\bibitem{KaonDecayChannel1} 
Kobayashi K \textit{et al.} 
2005
Search for nucleon decay via modes favored by supersymmetric grand unification models in Super-Kamiokande-I
\textit{Phys.~Rev.~}D 
\textbf{72} 
052007

\bibitem{KaonDecayChannel2}
Babu K S, Pati J C and Wilczek F 
1998
Suggested new modes in supersymmetric proton decay
\textit{Phys.~Lett.~}B 
\textbf{42} 
3337-47

\bibitem{APEX2}
Geer S H and Kennedy D C 
2000
A New Limit on the Antiproton Lifetime
\textit{Astrophys.~J.~}\textbf{53} 
2648-52


\bibitem{Thompson2016}
Goodwin F, Stutter G, Thompson R C and Segal D M 
2016
Resolved-Sideband Laser Cooling in a Penning Trap
\textit{Phys.~Rev.~Lett.~}\textbf{116}
143002

\bibitem{Wineland1990} 
Wineland D J and Heinzen D J 
1990
Quantum-limited cooling and detection of radio-frequency oscillations by laser-cooled ions
\textit{Phys.~Rev.~}A 
\textbf{42} 
2977-94

\bibitem{Jerry5poleTrap} 
Gabrielse G, Haarsma L, Rolston S L
1989 
Open-endcap Penning traps for high precision experiments
\textit{Int.~J.~Mass Spec.~}\textbf{88} 
319-32


\bibitem{JackElise2019}
Devlin J A \textit{et al.} 2019
A superconducting solenoid system with adjustable shielding factor
\textit{Phys.~Rev.~Appl.~} accepted

\bibitem{Derevianko}
Derevianko A and Pospelov M
2014
Hunting for topological dark matter with atomic clocks
\textit{Nature Physics}
\textbf{10} 
933-6

\bibitem{NuclSpinComag}
Wu T \textit{et al.} 2019
Search for Axionlike Dark Matter with a Liquid-State Nuclear Spin Comagnetometer
\textit{Phys.~Rev.~Lett.~}\textbf{122} 
191302

\bibitem{Tseng1993}
Tseng C H \textit{et al.} 
1993
Portable Trap Carries Particles 5000-Kilometers
\textit{Hyp.~Int.~}\textbf{76} 
81-93

\bibitem{Wada1993}
Wada M and Yamazaki Y 
2004 
Technical developments toward antiprotonic atoms for nuclear structure studies of radioactive nuclei
\textit{Nucl.~Instr.~Meth.~}B \textbf{214} 196-200

\bibitem{LewisMedical}
Lewis R A 
1997
Antiproton portable traps and medical applications
\textit{Hyp.~Int.~}\textbf{109} 
155-64

\bibitem{PUMA}
Aumann T \textit{et al.} 2019
Experiment Proposal for PUMA: antiprotons and radioactive nuclei
\textit{CERN Document Server} CERN-SPSC-2019-033 / SPSC-P-361



\bibitem{Reid} 
Reid R 
1975
Nuclear Magnetic Shielding in Hydrogen Molecule
\textit{Phys.~Rev.~}A 
\textbf{11} 
403-8

\bibitem{Grange} 
Grange J \textit{et al.}
2015
Muon (g-2) Technical Design Report
\textit{arXiv}:1501.06858

\bibitem{Fei} 
Fei X, Hughes V W and Prigl R
1997
Precision measurement of the magnetic field in terms of the free-proton NMR frequency
\textit{Nucl.~Instr.~Meth.~}A 
\textbf{394} 
349-56

\bibitem{Chupp} 
Chupp T and Swanson S
2001
Medical imaging with laser-polarized noble gases
\textit{Adv.~At.~Mol.~Opt.~Phy.~}\textbf{45} 41-98

\bibitem{Rudzinski} 
Rudzinski A, Puchalski M and Pachucki K
2009 
Relativistic, QED, and nuclear mass effects in the magnetic shielding of He-3
\textit{J.~Chem.~Phys.~}\textbf{130} 
244102 

\bibitem{Nikiel} 
Nikiel A, Bl\"umler P, Heil W, Hehn M, Karpuk S, Maul A, Otten E, Schreiber L M and Terekhov M
2014
Ultrasensitive $^3$He magnetometer for measurements of high magnetic fields
\textit{Eur.~Phys.~J.~}D 
\textbf{68} 
330

\bibitem{Flowers} 
Flowers J L, Petley B W and Richards M G
1993
\textit{Metrologia} 
A Measurement of the Nuclear Magnetic Moment of the He-3 Atom in terms of that of the Proton
\textbf{30} 
75-87

\bibitem{Jackowski} 
Jackowski K, Jaszunski M and Wilczek M
2010
Alternative Approach to the Standardization of NMR Spectra. Direct Measurement of Nuclear Magnetic Shielding in Molecules
\textit{J.~Phys.~Chem.~}A \textbf{114} 
2471-5

\bibitem{dehmelt} 
Schuessler H A, Fortson E N and Dehmelt H G 
1969 
Hyperfine Structure of Ground State of He-3+ by Ion-Storage Exchange-Collision Technique
\textit{Phys.~Rev.~}\textbf{187} 
5-38























\end{thebibliography}
\end{document}